\documentclass[aps,prb,twocolumn,showpacs,groupedaddress]{revtex4}
\usepackage{graphicx}

\begin{document}

\title{Ferroelectricity in perovskite $HoMnO_3$ and $YMnO_3$}
\author{Bernd Lorenz$^{1}$, Ya-Qi Wang$^{1}$, and Ching-Wu Chu$^{1,2,3}$}
\affiliation{$^{1}$Department of Physics and TCSUH, University of
Houston, Houston, TX 77204-5002} \affiliation{$^{2}$Lawrence
Berkeley National Laboratory, 1 Cyclotron Road, Berkeley, CA 94720}
\affiliation{$^{3}$Hong Kong University of Science and Technology,
Hong Kong, China}
\date{\today }

\begin{abstract}
% insert abstract here
Ferroelectricity is observed in orthorhombic $HoMnO_3$ and $YMnO_3$
at the magnetic lock-in transitions into an E-type structure or an
incommensurate phase with a temperature independent wave vector,
respectively. In $HoMnO_3$ the ferroelectric polarization strongly
depends on the external magnetic field indicating the involvement of
the rare earth moment order in this compound. The results are
discussed within the framework of recent theoretical models, in
particular the double exchange driven polar displacements predicted
for E-type magnetic structures.
\end{abstract}

 \maketitle

% insert suggested PACS numbers in braces on next line

% Use the \preprint command to place your local institutional report
% number in the upper righthand corner of the title page in preprint mode.
% Multiple \preprint commands are allowed.
% Use the 'preprintnumbers' class option to override journal defaults
% to display numbers if necessary
%\preprint{}

%Title of paper

% repeat the \author .. \affiliation  etc. as needed
% \email, \thanks, \homepage, \altaffiliation all apply to the current
% author. Explanatory text should go in the []'s, actual e-mail
% address or url should go in the {}'s for \email and \homepage.
% Please use the appropriate macro foreach each type of information

% \affiliation command applies to all authors since the last
% \affiliation command. The \affiliation command should follow the
% other information
% \affiliation can be followed by \email, \homepage, \thanks as well.

%\email[]{Your e-mail address}
%\homepage[]{Your web page}
%\thanks{}
%\altaffiliation{}

%Collaboration name if desired (requires use of superscriptaddress
%option in \documentclass). \noaffiliation is required (may also be
%used with the \author command).
%\collaboration can be followed by \email, \homepage, \thanks as well.
%\collaboration{}
%\noaffiliation

% insert suggested keywords - APS authors don't need to do this
%\keywords{}

%\maketitle must follow title, authors, abstract, \pacs, and \keywords

% body of paper here - Use proper section commands
% References should be done using the \cite, \ref, and \label commands

\section{Introduction}

Multiferroic magnetoelectric manganites have attracted increasing
attention because of the coexistence of ferroelectric and magnetic
orders, the control of polarization by external and internal
magnetic fields as well as by external pressure, the role of spin
frustration in stabilizing improper ferroelectricity, and the rich
phase diagrams with cascades of magnetic and ferroelectric phase
transitions upon changing temperature or magnetic
fields.\cite{fiebig:05,prellier:05} In hexagonal manganites,
$RMnO_3$ ($R=Y$, $Ho$ to $Lu$), ferroelectricity arises well above
the antiferromagnetic (AFM) ordering temperature and the coupling
between the ferroelectric (FE) order and the frustrated AFM
$Mn$-spin structure is indirect (the linear magnetoelectric effect
is forbidden by symmetry) and mediated by magnetoelastic
interactions.\cite{huang:97,lorenz:04,delacruz:05} In contrast, the
ferroelectricity in orthorhombic $TbMnO_3$ and the kagom\'{e}
staircase compound $Ni_3V_2O_8$ has been shown to be triggered by a
magnetic phase transition from a sinusoidal to a helical spin
density wave with the spatial inversion symmetry broken in the
latter phase.\cite{kenzelmann:05,lawes:05} The magnetoelastic
coupling leads to a distortion of the lattice with a net FE
polarization perpendicular to the propagation vector of the spin
density wave and the axis of rotation of the helical
modulation.\cite{mostovoy:06} This empirical picture is based on
symmetry considerations, it suggests a trilinear coupling of the
magnetic order parameters and the FE polarization, and it is
supported by a recent microscopic model for
$Ni_3V_2O_8$.\cite{harris:06} The orthorhombic $RMn_2O_5$ rare earth
manganites also show ferroelectricity at the magnetic lock-in
transition into a commensurate magnetic structure. The microscopic
origin of the FE order has yet to be understood, however, it was
suggested that the magnetic frustration among the Mn-spins is
reduced by a displacement of oxygen and manganese ions with a
macroscopic polarization along the orthorhombic
$b$-axis.\cite{kagomiya:03,delacruz:06} In a more detailed
investigation of the spin structure of $TbMn_2O_5$ and $YMn_2O_5$
Chapon et al. have explained the ferroelectricity in the
commensurate phase of $RMn_2O_5$ as induced by an acentric
spin-density wave.\cite{chapon:04,chapon:06}

The magnitude of the FE polarization in the above mentioned
manganites and vanadates is relatively small with typical values of
100 nC/cm$^2$ at low temperatures. This was attributed in part to
the small antisymmetric Dzyaloshinskii-Moriya spin-spin interaction
resulting from the spin-orbit coupling that plays a major role in
the ferroelectrics with helical magnetic
structures.\cite{sergienko:06} Alternative physical mechanisms
resulting in FE order induced by magnetic modulations with possibly
larger values of the FE polarization have therefore been searched
for. Sergienko et al.\cite{sergienko:06b} have recently proposed the
existence of FE order in the E-type magnetic structure of
orthorhombic $HoMnO_3$ and other compounds and predicted an orders
of magnitude larger polarization as compared to the manganites
discussed in the previous paragraph. The microscopic mechanism is a
competition between elastic energy and the energy gain due to the
(virtual) hopping of Mn-e$_g$ electrons along the ferromagnetically
aligned moments of an Mn-zigzag chain in the E-type magnetic
structure. Strong dielectric anomalies and an increase of the
dielectric constant of up to 60 \% at the magnetic transitions of
orthorhombic $HoMnO_3$ and $YMnO_3$ have indeed been reported by
us\cite{lorenz:04b} suggesting the possible existence of FE. We have
therefore decided to search for FE in orthorhombic $HoMnO_3$ and
$YMnO_3$ and assess the role of the rare earth element. We prove the
existence of a macroscopic FE polarization $P$ in orthorhombic
$HoMnO_3$ and $YMnO_3$ by measuring the pyroelectric current that
arises from a change of $P(T)$ during cooling or heating through the
magnetic and FE phase transitions.

The $RMnO_3$ manganites can be synthesized in the orthorhombic
structure for all rare earth ions, however, for $R=Ho$ to $Lu$ and
$Y$ the perovskite structure is metastable and the synthesis has to
be conducted under high-pressure
conditions.\cite{lorenz:04b,zhou:06} The magnetic structures of
$HoMnO_3$ and $YMnO_3$ have been investigated by neutron scattering.
Both compounds undergo a phase transition into an incommensurate
magnetic structure at $T_N\approx42$ K and at lower temperature the
magnetic order locks into a commensurate E-type order
($HoMnO_3$)\cite{munoz:01b} or into an incommensurate phase with a
temperature independent modulation vector ($YMnO_3$).\cite{munoz:02}
The stability of the E-type magnetic order in $HoMnO_3$ as the
ground state configuration was recently confirmed by first principle
calculations.\cite{picozzi:06} Dielectric anomalies are observed at
all magnetic phase transitions.\cite{lorenz:04b}

\section{Experimental}

The samples of $HoMnO_3$ and $YMnO_3$ used in this report have been
synthesized under high-pressure conditions as described
elsewhere.\cite{lorenz:04b} X-ray spectra indicate a pure
orthorhombic phase with no detectable reflections of impurity
phases. Polycrystalline pellets of high density have been shaped for
pyroelectric current measurements. The typical sample thickness was
0.4 mm and the contact area was of the order of 8 mm$^2$. The
Physical Property Measurement System (Quantum Design) was used for
temperature and magnetic field control. The pyroelectric current was
measured upon heating in zero electric field employing the Keithley
6517A electrometer after cooling the sample in electric fields of up
to 5 kV/cm and shortening the contacts at the lowest temperature for
15 min. The polarization was calculated by integrating the
pyroelectric current. Although the samples are polycrystalline
(single crystals of high-pressure synthesized samples are not
available) the current signal was clearly detected. The measurement
represents an average of the polarization over all possible grain
orientations and the data provide a lower limit for the actual FE
polarization that is supposed to be aligned with the $a$-axis (space
group $Pbnm$).\cite{sergienko:06b}

\section{Results and Discussion}

\begin{figure}
\begin{center}
\includegraphics[angle=-90,width=2.5in]{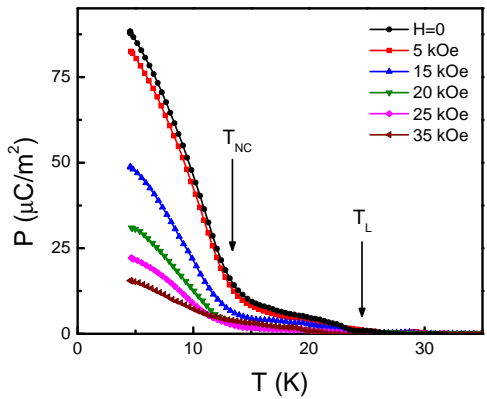}
\end{center}
\caption{(Color online) Temperature dependence of the ferroelectric
polarization $P$ of HoMnO$_3$ in different magnetic fields H
(increasing from top to bottom curve).}
\end{figure}

The temperature dependence of the FE polarization of $HoMnO_3$
calculated from the pyroelectric current data is shown in Fig. 1 at
different magnetic fields. The spontaneous polarization $\vec{P}$
starts to grow at $T_L$=26 K, the temperature of the lock-in
transition into the AFM E-phase. The neutron scattering
data\cite{munoz:01b} show that the magnetic moments of the Mn-ions
change continuously below $T_N$ in passing through $T_L$ and the
$Ho$-moment increases smoothly below $T_L$. The magnitude of
$\vec{P}$ is small, in contrast to the large value predicted by
Sergienko et al.,\cite{sergienko:06b} and it increases rapidly only
below 15 K, the temperature at which a major change in the magnetic
structure of the Ho-moments has been reported.\cite{munoz:01b}
According to the neutron scattering results the collinear AFM order
of the Ho-moments oriented along the $b$-axis sets in at $T_L$ with
the sublattice magnetization increasing significantly at lower $T$.
At $T_{NC}$=15 K the Ho-moments rotate in the $a$-$b$ plane and form
a noncollinear magnetic structure.\cite{munoz:01b} The polarization
(Fig. 1) exhibits a significant increase at this temperature
suggesting an involvement of the rare earth magnetic order in
stabilizing the FE displacements. The temperature dependence of $P$
is indeed qualitatively similar to that of the
$Ho$-moment.\cite{munoz:01b} The FE polarization is dramatically
reduced in external magnetic fields as shown in Figs. 1 and 2 up to
70 kOe. Increasing the magnetic field at a rate of 200 Oe/s at 4.5
K, after poling the FE domains during cooling, indeed generates a
current that proves the decrease of the FE polarization as shown in
Fig. 2. The inset of Fig. 2 displays the estimated polarization. In
agreement with the magnetic field dependence of the dielectric
constant\cite{lorenz:04b} $P(H)$ experiences the major decrease
above 10 kOe where a metamagnetic transition (alignment of the
$Ho$-moments with the field) had been
reported.\cite{munoz:01b,lorenz:04b}

\begin{figure}
\begin{center}
\includegraphics[angle=-90,width=2.5in]{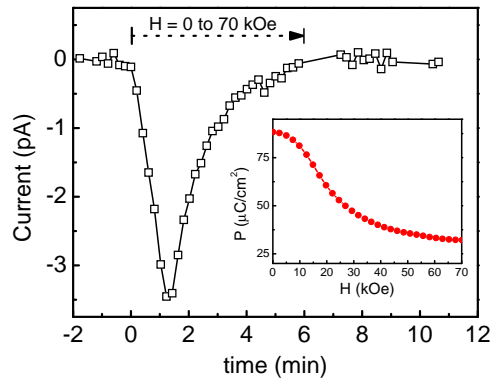}
\end{center}
\caption{(Color online) Decrease of FE polarization of $HoMnO_3$
with increasing magnetic field at 4.5 K. the main panel shows the
discharging current. Inset: Ferroelectric polarization $P(H)$.}
\end{figure}

The strong correlation of the polarization change below 15 K with
the $Ho$ magnetic order and its sensitivity to external magnetic
fields raises the question about the microscopic origin of the
ferroelectricity and the role of $Mn$-spins and rare earth magnetic
moments. According to the recent theoretical
proposal\cite{sergienko:06b} the rare earth moment should not be
essential for the FE order in the compound, but magnetoelastic
effects along the ferromagnetic zigzag chains of $Mn$-spins are
responsible for the FE distortions. We have, therefore, investigated
rare-earth free orthorhombic $YMnO_3$ with respect to
ferroelectricity. The pyroelectric current measured on a
polycrystalline sample indeed reveals a FE transition taking place
at 28 K. The measured polarization is shown in Fig. 3. Unlike the
$HoMnO_3$, the polarization of $YMnO_3$ does not depend on the
magnetic field up to 70 kOe. The FE polarization in $YMnO_3$ is
larger than that determined for $HoMnO_3$ and it arises exactly at
the lock-in temperature $T_L=28 K$ of the incommensurate $Mn$ spin
density wave. Neutron scattering experiments suggested that below
$T_L$ the spin modulation is still incommensurate with $q=0.434$
along the $b$-axis but it is locked into a temperature independent
value.\cite{munoz:02} The origin of ferroelectricity in orthorhombic
$YMnO_3$, however, is difficult to understand within this collinear
IC magnetic order scheme. This issue will be discussed in detail
below.

The value of the measured polarization is largely affected by the
polycrystalline nature of the high-pressure synthesized samples.
However, large single crystals of high pressure synthesized
orthorhombic $RMnO_3$ are not available and all experiments have to
be conducted using dense ceramic samples. While this is not a
principle problem and ferroelectric ceramics are well known and even
commercially available, a few cautious measures have to be applied
to avoid experimental artifacts affecting the results and
interpretations. The electrical current $i$ between the two
electrodes attached to a parallel plate dielectric can be expressed
as
\begin{equation}
i=C\frac{dV}{dt}+A\frac{dP}{dt}+\frac{V}{R}
\end{equation}
$V$ is the (time dependent) applied voltage,
$C=\varepsilon\varepsilon_0A/L$ is the capacitance, $A$ and $L$ are
the contact area and sample thickness, respectively, and $R$ is the
sample resistance. The three terms in equation (1) are the
capacitive, ferroelectric, and resistive currents. The samples of
$HoMnO_3$ and $YMnO_3$ investigated in this work are highly
resistive with a dc-resistance at low temperatures larger than
$10^9$ $\Omega$. This excludes any significant resistive currents
and it indicates the high quality of the compounds. In order to
separate the current that is due to a sole change of the
ferroelectric polarization all measurements have been conducted at
zero bias voltage with increasing temperature (thus excluding the
capacitive current) after cooling the samples in a constant electric
field. Before starting the pyroelectric current measurements at the
lowest temperature the electrodes have been shortened for up to 15
min in order to release any charges in the system. This procedure
excludes any artifact due to the change of large electric fields,
trapped charges, etc. as discussed earlier in the case of extremely
inhomogeneous systems (for example ferroelectric
copolymers).\cite{guy:91} The possibility that ferroelectricity
arises in the grain boundaries or the surface of the grains cannot
be completely ignored. However, it appears very unlikely that the
pyroelectric current measured between the two metallic electrodes is
solely generated by polarization charges in the grain boundary. The
total area of the grain boundaries underneath the electrode is small
as compared to the contact area and any contribution from a possible
polarization in the grain boundaries is expected to be negligible.

Due to the random orientation of the grains in the polycrystalline
sample the measured polarization has to be lower than the intrinsic
$P$ for a single grain or FE domain. First, the measurement can only
provide data for the component of $\vec{P}$ perpendicular to the
contact surface averaging over all possible grain orientations since
it is clear that the FE polarization is directed along one of the
principal crystallographic orientations. According to the theory of
Sergienko et al.\cite{sergienko:06b} $\vec{P}$ is aligned with the
orthorhombic $a$-axis. Secondly, cooling in strong electric field is
necessary to align the FE domains within a single grain. The
electric field component in the direction of the FE polarization,
however, depends on the grain orientation and it decreases if the
angle between the electric field and $\vec{P}$ increases. Therefore,
grains with a less perfect FE domain alignment will always exist
causing the macroscopic polarization to be smaller than that for a
properly oriented single crystal. In addition, even with a perfect
grain orientation the minimum electric field $E_{min}$ to align all
domains along the field direction it is not known and the maximum
applied field of 5 kV/cm in this experiment may be smaller than
$E_{min}$. Indeed, plotting the maximum $P$ of $YMnO_3$ at the
lowest temperature as a function of the poling electric field (Inset
to Fig. 3) reveals that the saturation polarization was not achieved
in the current experiment. Increasing the poling field further is
experimentally difficult due to the limited strength of the sample
and the environment with respect to a dielectric breakdown. Single
crystals of orthorhombic $HoMnO_3$ and $YMnO_3$ are highly desirable
but not yet available. Alternatively, the growth of thin epitaxial
films on appropriate substrates with the orientation of $\vec{P}$
perpendicular to the film surface will avoid the random grain
problem and could be used to determine the intrinsic values of the
FE polarization.

\begin{figure}
\begin{center}
\includegraphics[angle=-90,width=2.5in]{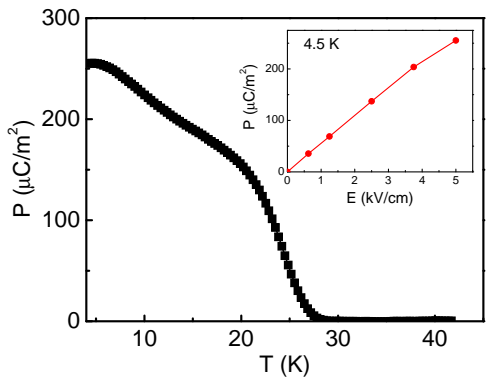}
\end{center}
\caption{(Color online) Temperature dependence of the ferroelectric
polarization $P$ of orthorhombic YMnO$_3$. The inset shows the
maximum polarization at 4.5 K for different poling fields.}
\end{figure}

\begin{figure}
\begin{center}
\includegraphics[angle=-90,width=3.5in]{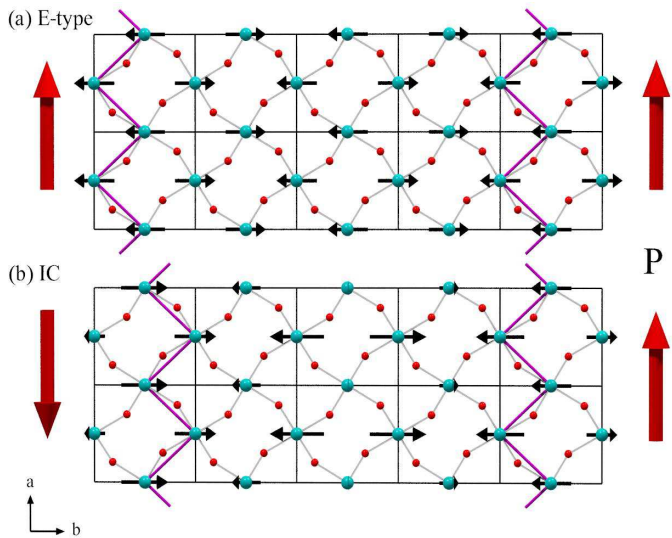}
\end{center}
\caption{(Color online) Magnetic order and ferroelectric
polarization arising from the double exchange mechanism for (a) the
E-type commensurate structure ($HoMnO_3$) and (b) the sinusoidal
incommensurate order proposed for $YMnO_3$.\cite{munoz:02} Large
spheres: Mn-ions with spins, small spheres: oxygen ions. The FM
aligned zigzag chains are shown for the first and last unit cells
and the local polarization in these chains is indicated by the large
vertical arrows.}
\end{figure}

The microscopic origin of the FE order in orthorhombic $RMnO_3$ is
still a matter of discussion. Several models have been proposed to
explain the ferroelectricity induced by magnetic orders. The phase
transition from sinusoidal spin modulation to a helical magnetic
structure was shown by symmetry arguments and Landau theory to break
the spatial inversion symmetry and allow for a macroscopic FE
polarization through a third order coupling between the magnetic
order parameter and the FE
polarization.\cite{kenzelmann:05,lawes:05,mostovoy:06} For the
orthorhombic $YMn_2O_5$ Chapon et al.\cite{chapon:06} have shown
that a non-centrosymmetric magnetic structure arises at the magnetic
lock-in transition within a system of AFM aligned zigzag chains of
$Mn^{4+}$- and $Mn^{3+}$-spins propagating along the $a$-axis with
an appropriate spin orientation and a phase factor. The FE
polarization than arises from a strong magnetoelastic coupling.
Evidence of sizable magnetoelastic effects has been found recently
in large anomalies of the thermal expansivities at the FE
transitions in $RMn_2O_5$.\cite{delacruz:06} The most recent
proposal involves the double exchange mechanism virtually coupling
the $e_g$-electrons of neighboring $Mn^{3+}$ ions with the same
direction of the magnetic moment that competes with the elastic
energy.\cite{sergienko:06b} In an E-type magnetic modulation with
ferromagnetic zigzag chains of $Mn$-spins this mechanism can
stabilize ferroelectricity through a cooperative displacement of
oxygen ions generating a macroscopic polarization along the
orthorhombic $a$-axis. This model could explain the observed
ferroelectricity in $HoMnO_3$.

Our data for $HoMnO_3$ shown in Fig. 1, however, indicate that the
case is more complex because of the involvement of the rare earth
magnetic moment. The $Ho$-moment apparently plays a crucial role and
is responsible for the rapid increase of the FE polarization below
15 K (Fig. 1). While in principle the FE displacement below $T_L$
can be explained by the double exchange
mechanism\cite{sergienko:06b} the details of the temperature
dependence of $P$ and its strong magnetic field dependence needs
further discussion. The rare earth moments in $HoMnO_3$ experience a
spin reorientation at $T_{NC}$=15 K\cite{munoz:01b} changing their
collinear alignment with the $b$-axis to a non-collinear magnetic
structure below $T_{NC}$. This non-collinear structure can give rise
to an additional FE displacement in a sense discussed by
Mostovoy\cite{mostovoy:06} and others\cite{kenzelmann:05,lawes:05}
through magnetoelastic interactions. The non-collinear $Ho$-moment
order below $T_{NC}$ would allow a FE displacement along the
$a$-axis (perpendicular to the propagation vector of the E-type
modulation and the rotation axis between the two spin arrangements)
which is incidentally the same direction as follows from the double
exchange mechanism involving the $Mn$-spin order. Therefore we
conclude that, in addition to the $Mn$-spin order, the rare earth
magnetic order plays a major role in stabilizing ferroelectricity in
$HoMnO_3$ at low temperatures. This is further supported by the
strong magnetic field dependence of the polarization in $HoMnO_3$
above 10 kOe similar to the one observed for the dielectric constant
data.\cite{lorenz:04b} The metamagnetic transition aligning the
$Ho$-moments with the field decreases the angle between the
$Ho$-moments. Since the polarization arising from the non-collinear
$Ho$-moment order is proportional to the vector product of the two
moments\cite{mostovoy:06} the FE polarization will decrease with the
decreasing angle between them.

The ferroelectricity observed in orthorhombic $YMnO_3$ (Fig. 3) is
more difficult to explain based on the published data for the
magnetic structure of the $Mn$-spins. Neutron scattering experiments
have been interpreted with contradicting results. Quezel et
al.\cite{quezel:74} proposed a helical $Mn$-spin order with an
antiparallel spin arrangement in the $a$-$c$ plane and an
incommensurate modulation along the $b$-axis with a relatively small
propagation vector ($q_y=0.079$). Although this spin structure
involves a helical modulation it cannot give rise to a FE
displacement within the Landau theory proposed by
Mostovoy\cite{mostovoy:06} and others since the propagation vector
and the rotation axis of the helical modulation are parallel and the
corresponding cross product of both vectors results to zero. Later
Munoz et al.\cite{munoz:02} revised the magnetic structure of
$YMnO_3$ and proposed an incommensurate sinusoidal spin order
propagating along the orthorhombic $b$-axis with the wave vector of
$q_y$=0.434. This magnetic structure is characterized by zigzag
chains of parallel spins in the $a$-$b$ plane directed along the
$a$-axis. The major difference with respect to the E-type
commensurate order is the modulation of the magnitude of the spins
in neighboring zigzag chains that follows the incommensurate wave.
In fact, the E-type spin structure follows for the commensurate
value $q_y$=0.5. The double exchange mechanism proposed by Sergienko
et al.\cite{sergienko:06b} does apply to a single FM zigzag chain
and may result in a local distortion along the chain direction with
a corresponding polarization. However, due to the incommensurate
modulation of the $Mn$-moments along the $b$-axis the folding
direction of the FM zigzag chains is reversed every few unit cells
resulting in a reversal of the local polarization in the
chain.\cite{sergienko:06c} The magnetic structure and the resulting
in-chain polarization is sketched in Fig. 4b and compared with the
E-type structure of Fig. 4a. The character of the dielectric order
in the case of $YMnO_3$ should be anti-ferroelectric with a periodic
modulation of the polarization according to the IC magnetic
modulation. The two polarizations shown at the right and left unit
cells in Fig. 4b are opposite in sign since the folding direction of
the two FM zigzag chains correspond to the two magnetic domains of
the commensurate E-type magnetic order.\cite{sergienko:06c} Within
the double exchange model the FE polarization does change sign if
the magnetic domain is reversed.\cite{sergienko:06b}

While the macroscopic FE polarization was clearly observed in our
experiments below the lock-in transition of orthorhombic $YMnO_3$
the physical mechanism leading to ferroelectricity is not clear. It
appears conceivable to assume that, similar to $TbMnO_3$ or
$HoMnO_3$, the ferroelectricity in $YMnO_3$ is improper and the
primary order parameter is of magnetic origin driving the FE
distortion through a polar magnetic order and the spin-lattice
coupling. Since the magnetic structures derived from both neutron
scattering investigations\cite{quezel:74,munoz:02} do not support a
macroscopic polarization based on the mechanisms proposed by
Mostovoy\cite{mostovoy:06} and Sergienko\cite{sergienko:06b} an
alternative theoretical description (for example by extending the
existing models or by developing new theories) or a reconsideration
of the magnetic structure seem to be necessary. In the phase diagram
of orthorhombic $RMnO_3$\cite{goto:04} the $YMnO_3$ is close to
$HoMnO_3$ on one side and $DyMnO_3$ and $TbMnO_3$ at the other side.
For $TbMnO_3$ the ferroelectricity was explained by the IC helical
magnetic structure of the $Mn$-spins as verified by neutron
scattering data.\cite{kenzelmann:05} High quality single crystals
are needed to resolve the magnetic structure in great detail and to
distinguish the non-collinear helical modulation from a collinear
sinusoidal spin arrangement. Unfortunately, single crystals of
metastable $RMnO_3$ compounds synthesized under high pressure
conditions are not available and neutron scattering experiments on
polycrystalline samples or powders may not reveal the exact
orientation of the spins. Therefore, it is possible that the
magnetic order in $YMnO_3$ is similar to that of $TbMnO_3$, locking
into a helical structure below $T_L$=28 K that would explain the
ferroelectricity below this temperature. Alternatively, since the
size of the $Y^{3+}$-ion is almost identical to the size of
$Ho^{3+}$ and the $Mn-O-Mn$ bond angle is very similar for both
compounds the E-type magnetic structure could be realized also in
$YMnO_3$ in which case the ferroelectricity can be explained by the
double exchange mechanism discussed by Sergienko et
al.\cite{sergienko:06b} However, this assumption is only valid if
the magnetic modulation below $T_L$ is commensurate (E-type), in
contrast to the most recent neutron scattering
results.\cite{munoz:02} Further and more detailed investigations of
the magnetic structure of $YMnO_3$ are necessary to resolve the
conflicts and to find an explanation for the observed
ferroelectricity in the compound.

\section{Summary}

We have proven the existence of ferroelectric order below the
magnetic lock-in transition in orthorhombic $HoMnO_3$ and $YMnO_3$.
The FE polarization detected in both compounds is relatively small.
$P$ values of other orthorhombic $RMnO_3$ and $RMn_2O_5$ are
distinctively larger. The value of 250 $\mu C/m^2$ of $YMnO_3$ is
more comparable to the FE polarization of $Ni_3V_2O_8$. Thus the
expectation of a two orders of magnitude higher polarization in the
E-type structure\cite{sergienko:06b} cannot be verified in the case
of $Ho/YMnO_3$. It is unlikely that the polycrystalline nature of
our samples and the factors discussed above can account for such a
dramatic reduction of the polarization. While large single crystals
of the compounds are not available it is suggested to investigate
the FE properties of thin films grown on appropriate substrates.

\begin{acknowledgments}
We like to thank I. Sergienko for stimulating discussions and for
communicating his work prior to publication. This work is supported
in part by the T.L.L. Temple Foundation, the J. J. and R. Moores
Endowment, and the State of Texas through TCSUH and at LBNL through
the US DOE, Contract No. DE-AC03-76SF00098.
\end{acknowledgments}

\bibliographystyle{phpf}
%\bibliography{HMO}

% Create the reference section using BibTeX:
%\bibliographystyle{plain}
%\bibliography{basename of .bib file}

\end{document}